\documentclass[a4paper,11pt]{article}

\usepackage{pos}
\usepackage{lipsum}
\usepackage{microtype}

\title{Spectrum of PeV Cosmic-Ray Protons and Helium Nuclei with IceCube}

\ShortTitle{Spectrum of PeV Cosmic-Ray Protons and Helium Nuclei with IceCube}

\author{The IceCube Collaboration \\{\normalsize \normalfont(a complete list of authors can be found at the end of the proceedings)}\\}

\emailAdd{saffer@kit.edu}

\abstract{
The IceCube Observatory comprises a cubic-kilometer particle detector deep in the Antarctic ice and the cosmic-ray air-shower array IceTop at the surface above. Previous analyses of the cosmic-ray composition have used coincident events with IceTop detecting the electromagnetic shower footprint as well as GeV muons, while the sensors submerged in the ice measure the TeV muons from the same events. The energy range of previous composition analyses, however, has been limited to 3\,PeV primary energy and above, whereas the IceTop all-particle energy spectrum has been extended down to 250\,TeV. This contribution presents a method to reconstruct the combined spectrum of cosmic-ray protons and helium nuclei, starting at 200\,TeV primary energy. The resulting H+He spectrum closes the gap in the measurements of light cosmic rays between IceCube as well as KASCADE and experiments measuring in the TeV energy range, such as DAMPE and HAWC.

\vspace{4mm}

{\bfseries Corresponding authors:}
Julian Saffer$^{1*}$\\
{$^{1}$ \itshape Institute of Experimental Particle Physics, Karlsruhe Institute of Technology}\\[4mm]
$^*$ Presenter
}

\ConferenceLogo{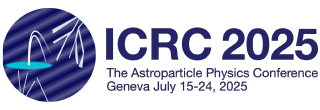}

\FullConference{39th International Cosmic Ray Conference (ICRC2025)\\
 15–24 July 2025\\
Geneva, Switzerland\\}

\begin{document}

\maketitle

\section{Introduction}

IceCube is an astroparticle telescope located at the geographic South Pole. The surface array IceTop covers an area of 1\,km$^2$ with 81 detector stations, each comprising a pair of ice-Cherenkov tanks, which are covered by a few meters of snow. Every tank houses two digital optical modules (DOMs). IceTop detects cosmic-ray-induced air showers in the primary energy range from a few hundred TeV to EeV. At the same time, a 1\,km$^3$ arrangement of 5160 DOMs at depths of 1450--2450\,m below IceTop registers the Cherenkov light emitted by muons in the ice, which either originate from neutrino interactions nearby or have been created as part of an air-shower's particle cascade.

As demonstrated in previous analyses~\cite{3year_paper, KoundalPlumSaffer_ICRC}, the selection and analysis of air-shower events that are first detected at the surface (by means of the electromagnetic component and GeV muons) and immediately after have their TeV muons seen deep in the ice allows for a reconstruction of both the primary cosmic ray's energy and its mass. This coincidence approach is possible thanks to the unique 3-dimensional multi-component setup of IceCube. Determining the composition of the hadronic cosmic rays is necessary to enhance the understanding of their production and propagation to Earth. In the past, composition studies with IceCube have been limited to a minimum primary energy of 3\,PeV. Direct detections of cosmic rays and air-shower observatories with a denser instrumentation but smaller total area, on the other hand, lose sensitivity beyond a few hundred TeV.

The IceTop all-particle spectrum has previously been extended down to 250\,TeV by triggering on small events that are contained in a central InFill region in the center of IceTop where the separation between stations is reduced~\cite{IceCube_lowenergy_CR}. In this work, the gap in the combined spectrum of protons and helium nuclei between direct and indirect measurements is closed by determining the fraction of events initiated by these light primaries and reconstructing their energy, reaching as low as 200\,TeV. The Monte Carlo (MC) simulations used for this analysis have been produced with CORSIKA~\cite{corsika} (version 7.7420), using the FLUKA interaction model~\cite{fluka} (version 2021.2) for lower-energy interactions below 80\,GeV as well as Sibyll~2.3d~\cite{Sibyll23d} as the high-energy hadronic interaction model. Results with alternative hadronic interaction models are obtained with QGSJet-II.04~\cite{QGSJETII04} and EPOS-LHC~\cite{EPOSLHC}.

\section{Event Selection}

The number of muons produced in an air shower depends on both the mass and the energy of the primary cosmic ray. Therefore, in order to identify the type of the primary particle, determining the muon multiplicity is not sufficient. Disentangling the energy-mass degeneracy also requires a simultaneous reconstruction of the primary energy. An air-shower event is recognized when at least three pairs of IceTop tanks (or two inside the InFill) are triggered simultaneously \cite{Aartsen:2016nxy}. Coincident events are selected, which provide the recorded charge deposits in IceTop DOMs caused predominantly by electromagnetic particles ($\mathrm{e}^\pm$, $\gamma$) close to the shower axis, as well as low-energy muons farther from the core. Furthermore, for coincident events the muon bundle energy loss can be estimated from the amount of Cherenkov light that is emitted by the high-energy muons in the deep detector. After a cleaning step, which reduces the amount of noise and background pulses, a hybrid shower reconstruction~\cite{RockBottom_ICRC2023, myTAUP, Saffer_PhDthesis} is applied to the IceTop pulses and the in-ice signal simultaneously. This algorithm fits a lateral distribution function to all IceTop pulses with its slope $\beta$ and shower size $S_{125}$ as two of the free parameters, which also include the core position as well as arrival time, direction and curvature of the shower front. Taking the in-ice pulses into account improves the directional accuracy due to the long lever arm of more than 1.5\,km. Apart from a converging reconstruction, the event selection cuts further demand the presence of in-ice signal in the top half of the detector, the containment of the reconstructed track within the borders of both IceTop and the in-ice detector as well as the energy proxy $S_{125}$ inside a specified range \cite{Saffer_PhDthesis}. Due to the geometry of IceCube, the zenith angle range of coincident events with the shower axis going through IceTop is limited to a maximum of about 30\textdegree.

\section{Light Cosmic-Ray Spectrum Reconstruction}

The task of reconstructing the primary energy and distinguishing light from heavy cosmic-ray primaries is given to a fully connected neural network. Provided with low-level event information like the lateral distribution of triggered IceTop stations and the total recorded charge of the muon-rich single-tank hits, as well as reconstructed event parameters such as shower size and in-ice energy loss, the network predicts the logarithmic primary energy and returns a score value between 0 and 1 quantifying the likelihood that the primary particle was a proton or helium nucleus rather than a heavier primary.

\begin{figure}
    \centering
    \begin{minipage}{.497\textwidth}
        \includegraphics[width=\textwidth]{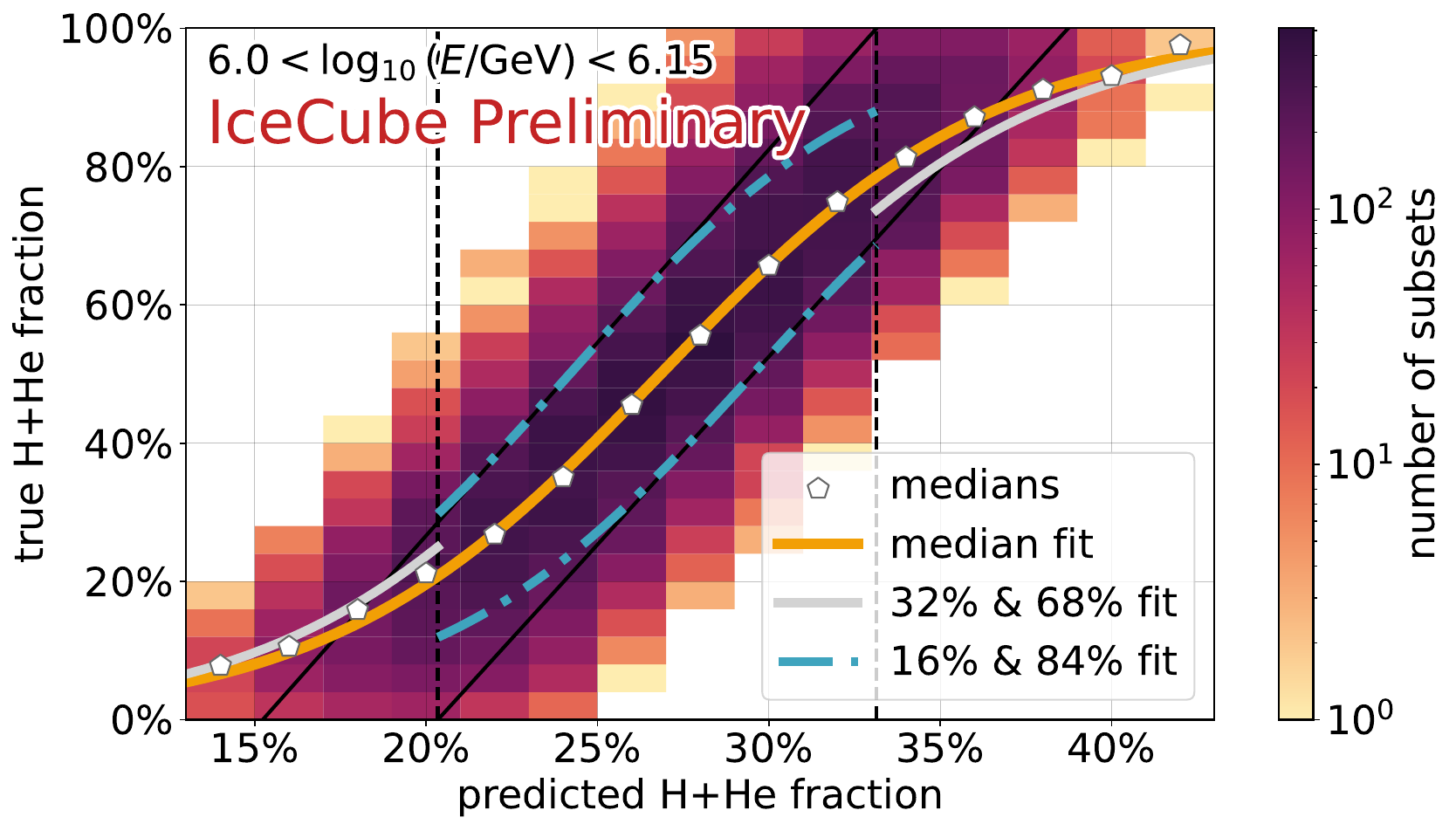}
    \end{minipage}
    \begin{minipage}{.497\textwidth}
        \includegraphics[width=\textwidth]{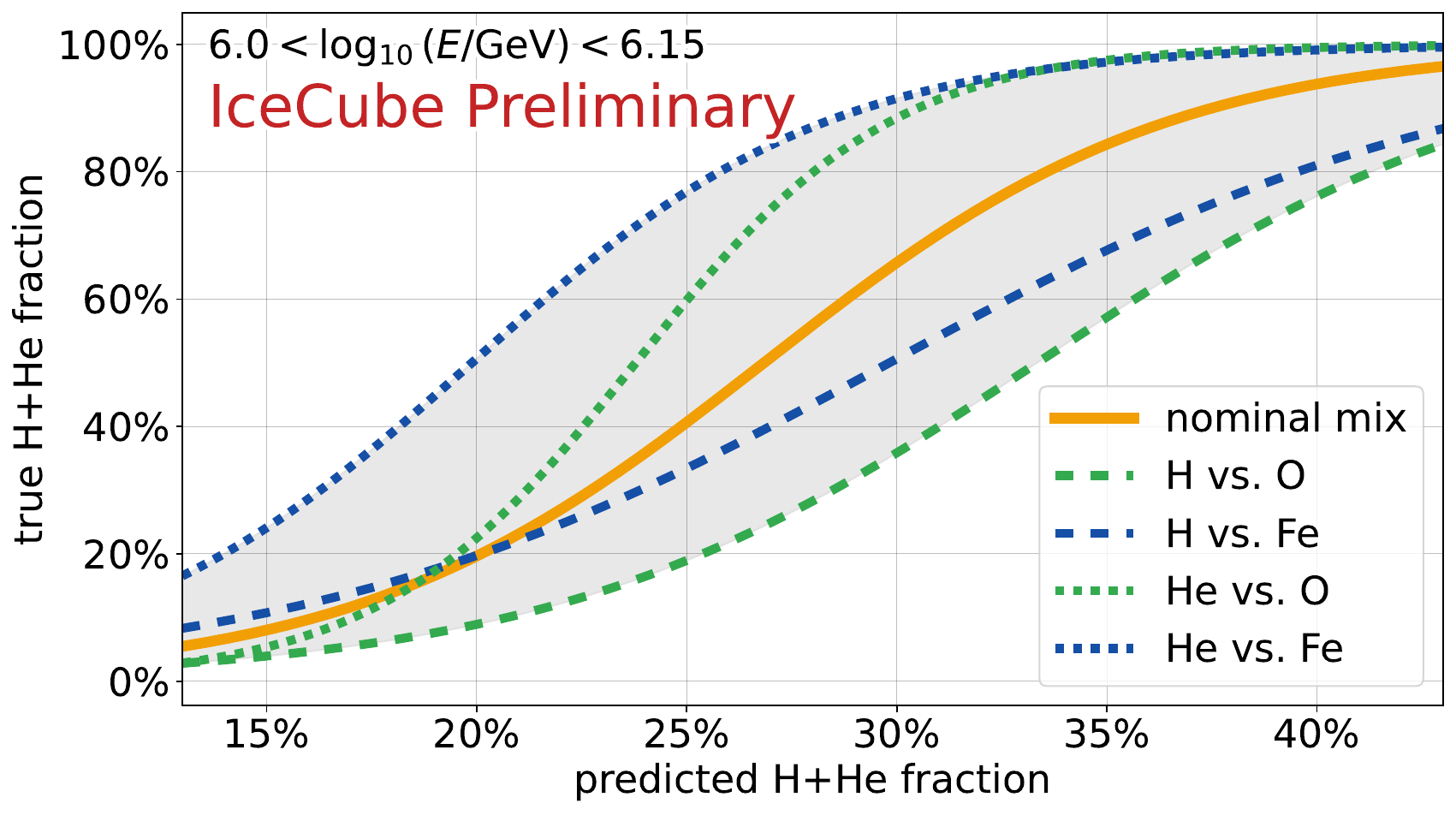}
    \end{minipage}
    \begin{minipage}{.497\textwidth}
        \caption{Calibration of the H+He fraction for the energy bin $10^6$--$10^{6.15}$\,GeV. \textit{Top left}: distribution of sampled subsets, nominal fit to the median true fraction and definition of sampling uncertainty via percentiles, \textit{top right}: composition uncertainty for the agnostic method (gray band) with extreme composition assumptions, \textit{bottom right}: composition uncertainty for the GSF-informed method (red band) using 200 variations of the best-fit GSF model.}
        \label{fig:light_fraction_calibration}
    \end{minipage}  
    \begin{minipage}{.497\textwidth}
        \includegraphics[width=\textwidth]{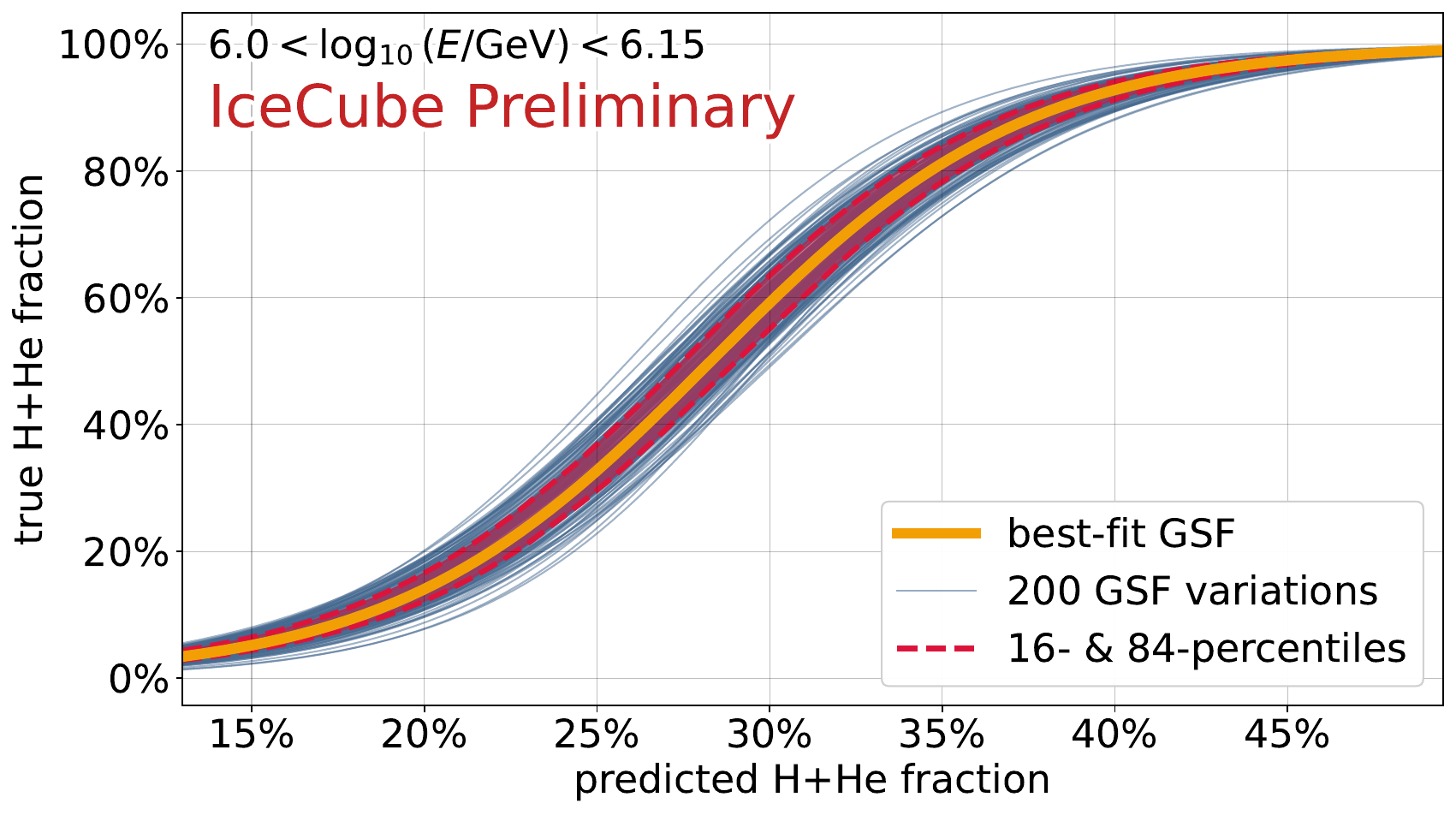}
    \end{minipage}  
\end{figure}

The all-particle bias, that is, the energy-binned median of $\log_{10}(E_\mathrm{pred}/E_\mathrm{MC})$ is determined and subtracted from the network's energy output for all events. The classification into events initiated by light and heavy primaries, on the other hand, requires a more complex handling. Since the ``lightness score'' distributions of light (proton, helium) and heavy (oxygen, iron) events overlap considerably, defining a classification threshold is accompanied by unavoidable contamination of the light sample with heavy events. 25\,330 different subsets of the simulated dataset are created and the actual fraction of proton- and helium-induced events is compared with the proportion of events with a score exceeding 0.35, which serves as a rough separation of light and heavy events. This is illustrated in Figure~\ref{fig:light_fraction_calibration} for a single energy bin. Even though the predicted light fractions systematically disagree with the true fractions, a correlation between both quantities is visible. The median true H+He fraction as a function of the network-predicted fraction follows a sigmoid shape, which is used to correct the estimated H+He fraction. The spread of true fraction in bins of predicted fraction is captured by fitting sigmoid functions to the 16th and 84th percentiles as well. The interval they span is considered to contain 68\% of subsets and is referred to as the systematic uncertainty due to sampling. When the predicted fraction is too low or too high, the distribution of true fractions becomes increasingly skewed. In that case, upper and lower limits on the true fraction are determined based on the 32nd and 68th percentiles, respectively, instead.

The slope and horizontal position of this calibration curve depend on the proton-to-helium ratio within the light flux component as well as the relation between the abundance of oxygen and iron events in the heavy part. The conservative way of defining systematic limits is to consider the four most extreme compositions, which combine one light and one heavy type each (top right in Fig.~\ref{fig:light_fraction_calibration}). This ``agnostic'' method is contrasted with an approach that is seeded with the GSF composition model~\cite{Dembinski_GSF}. The difference to the agnostic version is that the nominal calibration is based on the H/He ratio and O/Fe ratio as predicted by GSF rather than the equal mixture assumed in the agnostic case. Despite those relative composition assumptions in the light and heavy component respectively, the light/heavy ratio is not constrained. Furthermore, the systematic composition uncertainty is defined by the central 68\% of sigmoid fits, each corresponding to an individual variation of GSF, sampled using its covariance matrix (bottom right in Fig.~\ref{fig:light_fraction_calibration}).

In addition to sampling and composition uncertainty, statistical uncertainties on the efficiency, angular resolution, as well as variations of the measured event rate due to environmental effects (changes of atmospheric pressure and accumulation of snow) contribute to the total systematic uncertainty on the differential H+He flux. The uncertainty of Cherenkov light propagation in the ice is not included. A more detailed description of the analysis method can be found in Ref.~\cite{Saffer_PhDthesis}.

\section{Results}

\begin{figure}
    \centering
    \includegraphics[width=\textwidth]{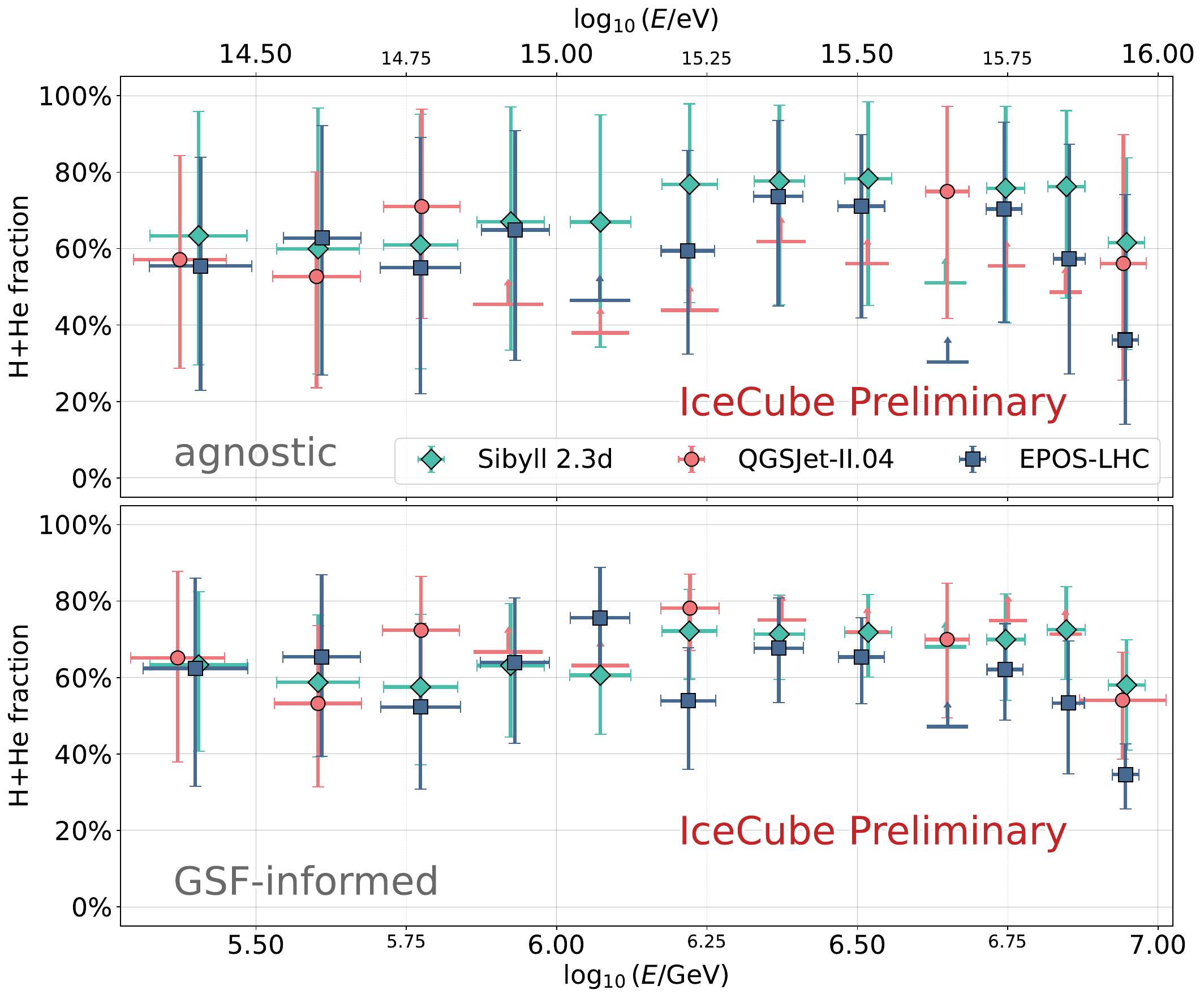}
    \caption{Reconstructed H+He fraction using the calibrations that are based on Monte Carlo produced with Sibyll~2.3d, QGSJet-II.04 and EPOS-LHC. Error bars represent the combined systematic uncertainty due to sampling as well as unknown composition in the vertical direction and energy resolution horizontally with the markers positioned at $\langle\log_{10}(E/\mathrm{GeV})\rangle$. \textit{Top}: agnostic calibration, \textit{bottom}: GSF-informed method.}
    \label{fig:light_fraction_hadr_models}
\end{figure}

The reconstruction and calibration are applied to one year of data, recorded by IceCube during one season from July 2018 to July 2019. In the PeV energy range, the light H+He component emerges as dominant (Figure~\ref{fig:light_fraction_hadr_models}). The fractions obtained using the calibrations based on the hadronic interaction models Sibyll~2.3d, QGSJet-II.04 and EPOS-LHC are mostly in agreement with each other within their systematic uncertainties. The heaviest composition is predicted by the approach based on EPOS-LHC simulations and beyond 8\,PeV the H+He fraction drops below 40\%. QGSJet-II.04, on the other hand, generally leads to a higher fraction of light cosmic-ray primaries compared to EPOS and Sibyll. However, for about half of the energy bins, the resulting fractions are defined by lower limits when using QGSJet. This is caused by the chosen definition of sampling uncertainty for high fractions, combined with a generally larger uncertainty due to lower Monte Carlo statistics compared with the calibration developed with Sibyll. Overall, the GSF-informed method yields a significantly reduced uncertainty compared to the agnostic approach.

\begin{figure}
    \centering
    \includegraphics[width=\textwidth]{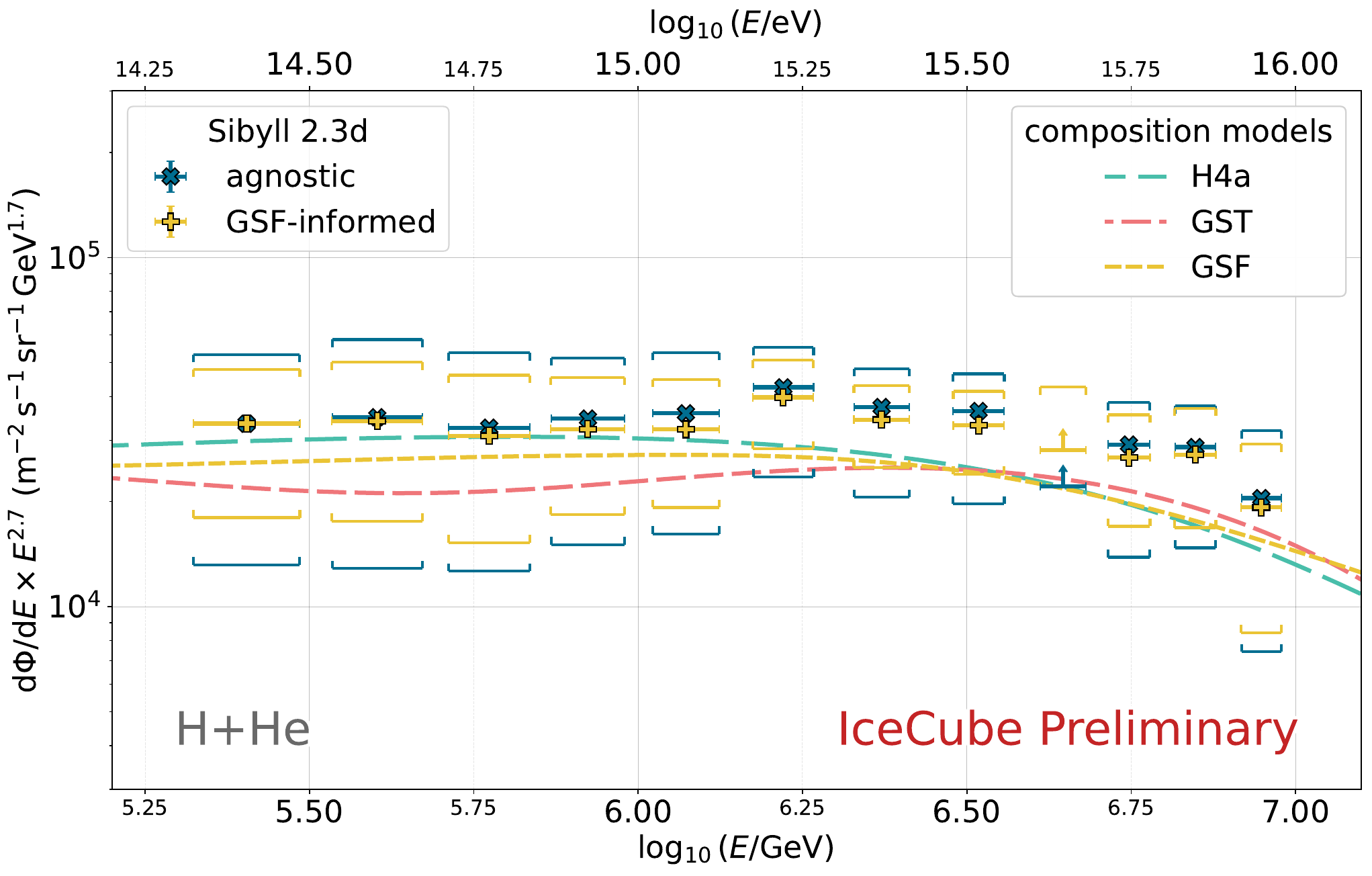}
    \caption{Compilation of the reconstructed H+He flux using the agnostic (blue) and the GSF-informed method (yellow), based on Sibyll~2.3d Monte Carlo. Systematic uncertainties are represented by brackets while error bars describe statistical uncertainty and energy resolution. The combined proton and helium flux of the composition models H4a, GST and GSF (dashed lines) is included for comparison.}
    \label{fig:light_flux_compo_models}
\end{figure}

Figure~\ref{fig:light_flux_compo_models} demonstrates that the reconstructed light flux obtained with the GSF-informed approach and its total uncertainty is contained entirely within the systematic uncertainty interval of the conservative agnostic result. Comparing the result with established flux models reveals good agreement up to about 1.5\,PeV ($\approx10^{6.2}\,\mathrm{GeV}$). Both, the agnostic and the GSF-informed method yield a higher flux than stated by the cosmic-ray flux models GSF~\cite{Dembinski_GSF}, H4a~\cite{Gaisser_H4a} and GST~\cite{Gaisser_GST}. With increasing primary energy, the reconstructed flux implies a continuation of the power-law spectrum with a spectral index of approximately -2.7 until it softens at about 5\,PeV ($\approx10^{6.7}\,\mathrm{GeV}$). Tension of the GSF-seeded result with the flux derived from the GST flux model is observed in the energy bin $10^{6.15}$--$10^{6.3}\,\mathrm{GeV}$. Even though the reconstructed H+He flux is systematically higher than predicted by the most recent composition models, the spectral shape resembles the one stated by GST. The slight dip around 600\,TeV ($\approx10^{5.8}\,\mathrm{GeV}$) matches the transition from the first to the second Galactic population described in Ref.~\cite{Gaisser_GST}.

\begin{figure}
    \centering
    \includegraphics[width=\textwidth]{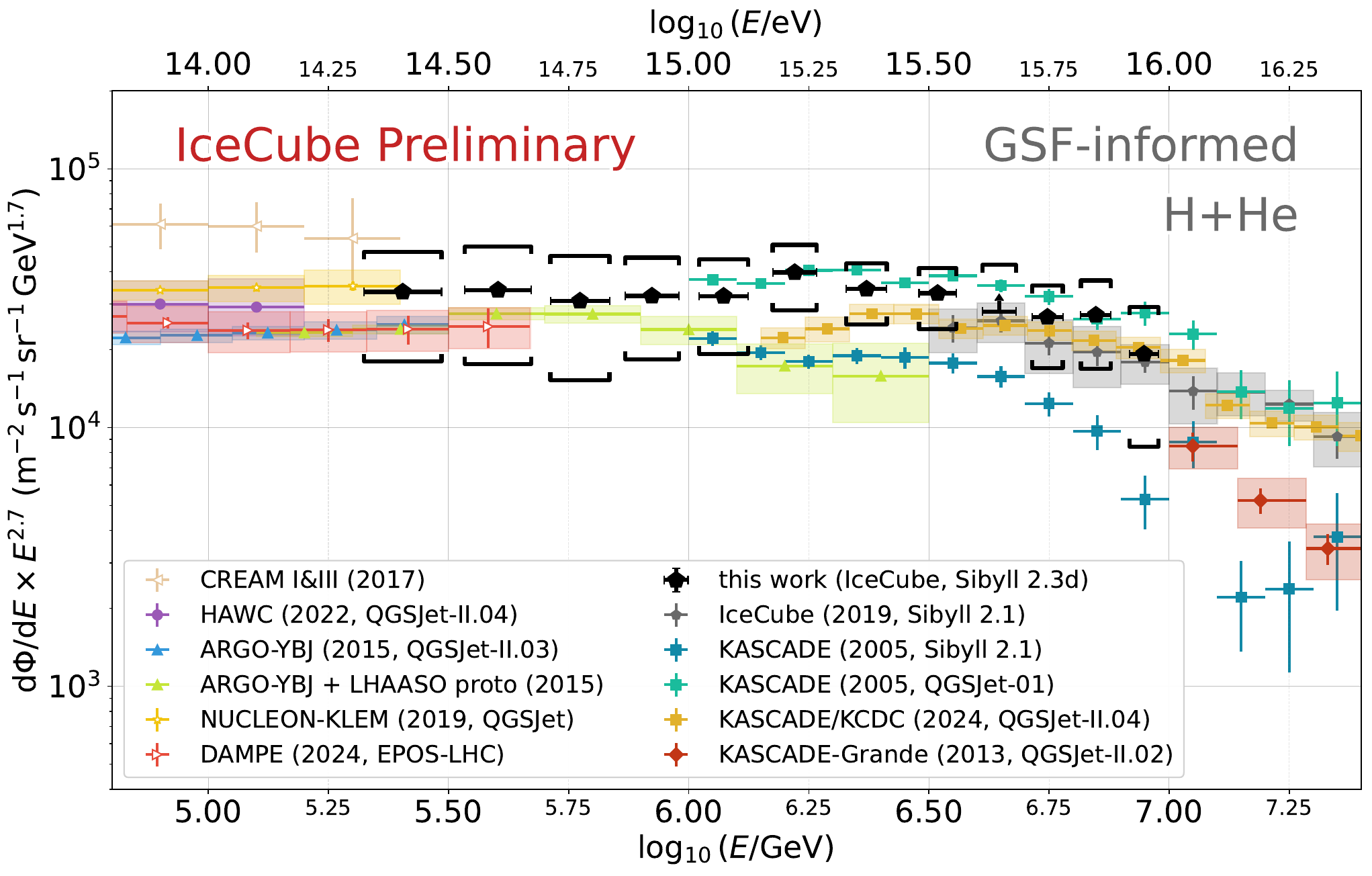}
    \caption{The differential spectrum of cosmic-ray protons and helium nuclei with one year of IceCube data (black). The hadronic model used is Sibyll~2.3d and the GSF-informed calibration method has been applied. Error bars represent energy resolution and statistical uncertainty in the horizontal and vertical direction, respectively. The total systematic uncertainty is indicated by brackets. The combined proton and helium flux as measured in the past by IceCube and other experiments is shown in various colors. Boxes represent their systematic and error bars their statistical uncertainties with the exception of CREAM and KASCADE (2005) where only total uncertainties are available. Open markers represent results obtained via direct detection. For references, see the text.}
    \label{fig:light_flux_other_experiments}
\end{figure}

A comparison with the light primary spectra from various other experiments and a previous IceCube result is shown in Figure~\ref{fig:light_flux_other_experiments}. Combined spectra of protons and helium are provided by DAMPE~\cite{DAMPE_pHe}, HAWC~\cite{HAWC_pHe}, ARGO-YBJ and LHAASO~\cite{ARGO-YBJ_pHe, ARGO-YBJ_LHAASO_pHe} whereas the separate spectra of protons and helium nuclei have been added in the case of CREAM~\cite{CREAM_pHe}, NUCLEON-KLEM~\cite{NUCLEON_KLEM_pHe}, KASCADE~\cite{KASCADE_pHe, KASCADE_ML} and KASCADE-Grande~\cite{KASCADE_Grande_pHe}. The sum of the fluxes of protons and helium measured by IceCube~\cite{3year_paper} is corrected for the estimated correlation between those two groups. Below 1\,PeV, the result of this analysis is in good agreement with direct experiments and smaller air-shower arrays. Interestingly, the variation among direct measurements by CREAM, NUCLEON-KLEM and DAMPE is larger than among the air-shower experiments HAWC and ARGO-YBJ. In the energy range 1--10\,PeV the new IceCube result that has been obtained with Sibyll~2.3d agrees well with the QGSJet flux from KASCADE as well as the previous IceCube spectrum, which is based on Sibyll~2.1. However, tension is observed with the Sibyll~2.1 measurement of the H+He flux at KASCADE, although the spectral shape agrees with the new IceCube result. The H+He knee just below 1\,PeV as declared by ARGO-YBJ cannot be confirmed.

\section{Conclusion and Outlook}

A new method for reconstructing the flux of protons and helium nuclei in the PeV energy range has been developed and applied to one year of data taken by IceCube. It involves a hybrid reconstruction of events that are detected by IceTop and the in-ice array in coincidence, followed by a neural network, which predicts energy and lightness of the primary cosmic ray. The subsequent calibration of the fraction of proton and helium events has been conducted in two ways: a conservative method with neutral assumptions and a version, which is seeded with the GSF composition model.

The calibrated H+He fractions obtained with the hadronic models Sibyll~2.3d, QGSJet-II.04 and EPOS-LHC agree within their uncertainty despite systematic differences. The dominance of the light component in the energy range around and below the knee could be confirmed. The H+He flux follows an $E^{-2.7}$ power law in first approximation. However, the spectral shape resembles that of the GST model. A knee of the light primaries appears at an energy slightly above 8\,PeV. No significant tensions are observed among either hadronic interaction models or with composition models. This is the first measurement of the H+He flux in the energy range from 300\,TeV to 1.4\,PeV with post-LHC models, closing the gap between direct and indirect measurements.

Since statistical uncertainties are negligible due to the high flux of cosmic rays around the knee, extending the analysis to more data has a low priority. However, for future enhancements of this study, it is imperative to reduce systematic uncertainties. Template fits might be a good alternative for estimating the H+He fraction that does not rely on further calibration. The environmental effect on the reconstructed flux could be decreased by improving the pressure correction of event rates and simulating a variety of South Pole atmospheres in CORSIKA. The systematic studies may be extended by investigating the effect of ice models on the propagation of light and hence the reconstructed composition. Higher sensitivity to low-energy air showers is expected with the completion of the IceCube surface array enhancement, which adds scintillators and radio antennas to the detector. Furthermore, the simulation database could be extended with event simulations using the latest updates to the Sibyll, QGSJet and EPOS hadronic interaction models.

\small
\bibliographystyle{ICRC}
\bibliography{references}

\clearpage

\section*{Full Author List: IceCube Collaboration}

\scriptsize
\noindent
R.~Abbasi$^{16}$,
M.~Ackermann$^{63}$,
J.~Adams$^{17}$,
S.~K.~Agarwalla$^{39,\: {\rm a}}$,
J.~A.~Aguilar$^{10}$,
M.~Ahlers$^{21}$,
J.M.~Alameddine$^{22}$,
S.~Ali$^{35}$,
N.~M.~Amin$^{43}$,
K.~Andeen$^{41}$
% ...
\\
\\
% ...
$^{10}$ Universit{\'e} Libre de Bruxelles, Science Faculty CP230, B-1050 Brussels, Belgium \\
% ...
$^{16}$ Department of Physics, Loyola University Chicago, Chicago, IL 60660, USA \\
$^{17}$ Dept. of Physics and Astronomy, University of Canterbury, Private Bag 4800, Christchurch, New Zealand \\
% ...
$^{21}$ Niels Bohr Institute, University of Copenhagen, DK-2100 Copenhagen, Denmark \\
$^{22}$ Dept. of Physics, TU Dortmund University, D-44221 Dortmund, Germany \\
% ...
$^{35}$ Dept. of Physics and Astronomy, University of Kansas, Lawrence, KS 66045, USA \\
% ...
$^{39}$ Dept. of Physics and Wisconsin IceCube Particle Astrophysics Center, University of Wisconsin{\textemdash}Madison, Madison, WI 53706, USA \\
% ...
$^{41}$ Department of Physics, Marquette University, Milwaukee, WI 53201, USA \\
% ...
$^{43}$ Bartol Research Institute and Dept. of Physics and Astronomy, University of Delaware, Newark, DE 19716, USA \\
% ...
$^{63}$ Deutsches Elektronen-Synchrotron DESY, Platanenallee 6, D-15738 Zeuthen, Germany \\
$^{\rm a}$ also at Institute of Physics, Sachivalaya Marg, Sainik School Post, Bhubaneswar 751005, India \\
% ...

\subsection*{Acknowledgments}

\noindent
The authors gratefully acknowledge the support from the following agencies and institutions:
USA {\textendash} U.S. National Science Foundation-Office of Polar Programs,
U.S. National Science Foundation-Physics Division,
U.S. National Science Foundation-EPSCoR,
U.S. National Science Foundation-Office of Advanced Cyberinfrastructure,
Wisconsin Alumni Research Foundation,
Center for High Throughput Computing (CHTC) at the University of Wisconsin{\textendash}Madison,
Open Science Grid (OSG),
Partnership to Advance Throughput Computing (PATh),
Advanced Cyberinfrastructure Coordination Ecosystem: Services {\&} Support (ACCESS),
Frontera and Ranch computing project at the Texas Advanced Computing Center,
U.S. Department of Energy-National Energy Research Scientific Computing Center,
Particle astrophysics research computing center at the University of Maryland,
Institute for Cyber-Enabled Research at Michigan State University,
Astroparticle physics computational facility at Marquette University,
NVIDIA Corporation,
and Google Cloud Platform;
Belgium {\textendash} Funds for Scientific Research (FRS-FNRS and FWO),
FWO Odysseus and Big Science programmes,
and Belgian Federal Science Policy Office (Belspo);
Germany {\textendash} Bundesministerium f{\"u}r Forschung, Technologie und Raumfahrt (BMFTR),
Deutsche Forschungsgemeinschaft (DFG),
Helmholtz Alliance for Astroparticle Physics (HAP),
Initiative and Networking Fund of the Helmholtz Association,
Deutsches Elektronen Synchrotron (DESY),
and High Performance Computing cluster of the RWTH Aachen;
Sweden {\textendash} Swedish Research Council,
Swedish Polar Research Secretariat,
Swedish National Infrastructure for Computing (SNIC),
and Knut and Alice Wallenberg Foundation;
European Union {\textendash} EGI Advanced Computing for research;
Australia {\textendash} Australian Research Council;
Canada {\textendash} Natural Sciences and Engineering Research Council of Canada,
Calcul Qu{\'e}bec, Compute Ontario, Canada Foundation for Innovation, WestGrid, and Digital Research Alliance of Canada;
Denmark {\textendash} Villum Fonden, Carlsberg Foundation, and European Commission;
New Zealand {\textendash} Marsden Fund;
Japan {\textendash} Japan Society for Promotion of Science (JSPS)
and Institute for Global Prominent Research (IGPR) of Chiba University;
Korea {\textendash} National Research Foundation of Korea (NRF);
Switzerland {\textendash} Swiss National Science Foundation (SNSF).

\bigskip
\noindent The authors gratefully acknowledge the computing time provided on the high-performance computer HoreKa by the National High-Performance Computing Center at KIT (NHR@KIT). This center is jointly supported by the Federal Ministry of Education and Research and the Ministry of Science, Research and the Arts of Baden-Württemberg, as part of the National High-Performance Computing (NHR) joint funding program (https://www.nhr-verein.de/en/our-partners). HoreKa is partly funded by the German Research Foundation (DFG).

\end{document}